\documentclass[aps,pra,twocolumn,showpacs]{revtex4-2}%

\usepackage{amsfonts}
\usepackage{amsmath}
\usepackage{amssymb}
\usepackage{graphicx}
\usepackage{dcolumn}
\usepackage{bm}

\begin{document}

\title{Temperature dependence of London penetration depth anisotropy\\ in superconductors with anisotropic order parameters}

 \author{ V. G. Kogan }
\email{kogan@ameslab.gov}
\affiliation{Ames Laboratory--DOE, Ames, Iowa 50011, USA }

 \author{R. Prozorov}
\email{prozorov@ameslab.gov}
 \affiliation{Ames Laboratory--DOE, Ames, IA 50011, USA}
 \affiliation{Department of Physics and Astronomy, Iowa State University, Ames, IA 50011, USA}

 \date{ \today}

\begin{abstract}
 We study effects of anisotropic order parameters on the  temperature dependence of London penetration depth anisotropy $\gamma_\lambda(T)$. After MgB$_2$, this dependence is commonly attributed to distinct gaps on multi-band   Fermi surfaces in superconductors.   We have found, however, that   the anisotropy parameter may depend  on temperature also in one-band materials with anisotropic order parameters $\Delta(T,\bm k_F)$,    a few such examples are given. We have found also that for different order parameters, the   temperature dependence of $\Delta(T)/\Delta(0)$ can be represented with good accuracy by  the  interpolation suggested by  D. Einzel, J.  Low Temp. Phys,  {\bf 131}, 1  (2003), which simplifies considerably the evaluation of $\gamma_\lambda(T)$. Of a particular interest is mixed order parameters of two symmetries for which $\gamma_\lambda(T)$ may go through a maximum for certain relative weight of two phases. Also, for this case we find that the ratio  $\Delta_{max}(0)/T_c$ may exceed substantially the weak coupling limit of 1.76. It, however, does not imply a strong coupling, rather it is due to significantly anisotropic angular variation of $\Delta$.
\end{abstract}
 \date{\today}
\maketitle

\section{Introduction}

The London penetration depth $\lambda$ is one of the major characteristics of superconductors. Most of materials studied  nowadays are anisotropic   with complicated Fermi surfaces and non-trivial order parameters $\Delta(\bm k)$ ($\bm k$ is the Fermi momentum). As a result,
$\lambda$ is also anisotropic; in uniaxial materials of interest here the $\lambda$ anisotropy is characterized by the anisotropy parameter $\gamma_\lambda=\lambda_c/\lambda_a$ ($a$ and $c$ stand for principal crystal directions). For a long time $\gamma_\lambda$ has been considered as a temperature independent constant. With the discovery of MgB$_2$ \cite{MgB2} it was found that $\gamma_\lambda(T)$ increases on warming \cite{Carrington} due to two different gaps on two groups of Fermi surface sheets \cite{Choi,K2002}. Since then, if a $T$ dependence of  $\gamma_\lambda$ is observed, it is commonly attributed to a multi-gap type of superconductivity.
  We show  below that, in fact, $\gamma_\lambda$ depends on $T$ also in  one-band case if the order parameter $\Delta$ is anisotropic even on isotropic Fermi surfaces.

We focus on the clean limit for two major reasons. Commonly after discovery of a new superconductor, an effort is made to obtain as clean single crystals as possible since  those are  better   to study the underlying physics.   Besides, in general, the scattering suppresses the anisotropy of $\lambda$, the  quantity of interest in this work.

Although our formal results are written in the form applicable to any Fermi surfaces, we consider only  Fermi spheres to separate effects of the order parameter symmetry on the anisotropy of $ \lambda$  from the effects of anisotropic Fermi surfaces.
 Another reason is experimental: there are materials currently  studied with nearly isotropic upper critical field, but with unusual non-monotonic $\gamma_\lambda(T)$ \cite{Kyuil}.

To our knowledge, up to now,   theoretical work on the temperature dependence of $\gamma_\lambda$ has been focused on  evaluation of $\gamma_\lambda $   at $T=0$ and  $T_c$ \cite{Kosh}. Assuming monotonic behavior of $\gamma_\lambda(T)$, the knowledge of the $\gamma_\lambda $ at the end points suffices for a qualitative description of this dependence.
 This assumption, however, is challenged by recent data on non-monotonic $\gamma_\lambda(T)$   \cite{Kyuil}.

  To evaluate the temperature dependence of the penetration depth and its anisotropy, one first has to calculate the equilibrium order parameter $\Delta(T)$, a non-trivial and time consuming task because one has to solve the self-consistency equation  of the theory (the gap equation). Instead, one can employ a version of the  interpolation scheme of D. Einzel \cite{Peter,Peter2,Einzel}, which provides an accurate representation of the BCS gap dependence $\Delta(T)$ for various order parameter symmetries. Moreover, we show that, in fact, the reduced $\Delta(T)/\Delta(0)$ as a function of reduced temperature $t=T/T_c$ has a nearly universal form for all order parameters we tested. This simplifies remarkably the task of evaluating $\Delta(T)$. We note, however, that all temperature-dependent results shown after Fig.1 were obtained using numerically exact solutions of the self-consistency Eq.(8).

\section{Approach}

Weak coupling  superconductors are described by a system of quasi-classical Eilenberger
equations \cite{E}.
For a clean material in the field absence, the Eilenberger functions $f,g$ satisfy \cite{E}:
\begin{eqnarray}
0&=&    \Delta\, g -  \omega f \,,\label{Eil1}\\
 1&=&g^2+f^2 \,.\,\qquad
\label{Eil3}
\label{normalization}
\end{eqnarray}
  Here,   $\Delta$ is the superconducting order parameter which might depend on the position at the Fermi surface, $ \omega=\pi T(2 n+1) $ are Matsubara frequencies; hereafter $\hbar=1$ and $k_B=1$.       This system yields:
\begin{eqnarray}
 f=  \Delta/\sqrt{ \Delta^2+    \omega^2}\,,\quad g=   \omega/ \sqrt{ \Delta^2+    \omega^2}\, .
\label{fg}
\end{eqnarray}
All equilibrium properties of uniform superconductors can be expressed in terms of $f$ and $g$.

Within  the separable model \cite{MarkKad}, the coupling responsible for superconductivity is assumed to have the form $V(\bm k,\bm k^\prime)=V_0 \Omega(\bm k)\Omega(\bm k^\prime)$, that leads to
\begin{eqnarray}
  \Delta(T,\bm k)=\Psi(T)\Omega(\bm k)\,.
\label{Del}
\end{eqnarray}
The function $\Omega(\bm k)$ is normalized \cite{Pokr}:
\begin{eqnarray}
  \langle \Omega^2\rangle =1\,,
\label{<Om2>}
\end{eqnarray}
  $\langle...\rangle$ stands for averaging over the Fermi surface.
This normalization is convenient, enough to mention the condensation energy at $T=0$ \cite{anis-criteria}:
 \begin{eqnarray}
F(0)=\frac{N(0)}{2}\langle\Delta^2(0)\rangle = \frac{N(0)}{2}\Psi^2(0)\,.
\label{F(0)}
\end{eqnarray}
where $N(0)$ is the density of states per spin.

The self-consistency equation which provides the temperature dependent order parameter $\Psi(T)$ reads \cite{K2002}:
\begin{eqnarray}
 \frac{\Psi}{2\pi T} \ln \frac{T_{c}}{T}=\sum_{\omega> 0}
\left (\frac{\Psi}{ \omega}- \Big\langle\Omega f\Big\rangle\right )\,,
\label{self-cons}
\end{eqnarray}
with $T_{c}$ being the critical temperature.
 The dimensionless form of this  equation is:
\begin{eqnarray}
 -\ln t =\sum_{n=0}^\infty \left(\frac{1}{n+1/2} - \left\langle \frac{\Omega^2}{\sqrt{(n+1/2)^2+\Omega^2\delta^2/t ^2}} \right\rangle \right).
  \nonumber\\
\label{self-onsc-clean}
\end{eqnarray}
where $n$ is Matsubara integer, $t=T/T_c$, and  $\delta=\Psi/2\pi T_c$.
Clearly, the solution $\delta(t)$ depends on  anisotropy of the  order parameter given by $ \Omega$.

     In particular, one obtains \cite{Peter2,anis-criteria}, see also  Appendix A:
         \begin{eqnarray}
 \frac{\Psi(0)} {T_c} = \frac{\pi} { e^{ \gamma}}\,e^{-\langle\Omega^2\ln|\Omega |\rangle}.
 \label{NV}
\end{eqnarray}
If $T\to T_c$, Eq.\,(\ref{self-onsc-clean}) yields:
         \begin{eqnarray}
 \Psi^2(T) = \frac{8\pi^2T_c^2} { 7\zeta(3) \,\langle\Omega^4 \rangle}\left(1-\frac{T}{T_c}\right).
  \label{GL-OP}
\end{eqnarray}


D. Einzel  constructed a remarkably good approximation to the $T$ dependence of the order parameter \cite{Peter,Peter2,Einzel}:
         \begin{eqnarray}
 \frac{\Psi(t)} {\Psi(0)} &=&
\tanh\left( \frac{T_c}{\Psi(0)}  \sqrt{\frac{8\pi^2(1-t)}{7\zeta(3)\langle\Omega^4\rangle\,t} }
 \right)\,.
  \label{Einzel}
\end{eqnarray}
  A more accurate interpolation can be constructed by including   terms of the order $(1-t)^2$  \cite{Einzel}. 

   For $t\to 1$ we readily obtain   Eq.\,(\ref{GL-OP}).
If $t\to 0$, $\Psi(t)\to\Psi(0)$ and deviates from $ \Psi(0)$ exponentially slow due to $\tanh$-function. 

At low temperatures, one uses $\tanh x \approx 1-2e^{-2 x}$ to obtain from Eq.\,(\ref{Einzel}):
         \begin{eqnarray}
 \frac{\Psi(t)} {\Psi(0)} = 1 - 2  \exp\left(-\frac{T_c}{\Psi(0)}
\sqrt{\frac{8\pi^2 }{7\zeta(3)\langle\Omega^4\rangle\,t} }
 \right)\,.
  \label{large x}
\end{eqnarray}
This differs from the BCS result
        \begin{eqnarray}
 \frac{\Delta(t)} {\Delta(0)} = 1 -  \sqrt{\frac{2\pi T }{\Delta(0)} }\, e^{ -\Delta(0)/T}
\,.
  \label{BCSlarge}
\end{eqnarray}
Although Eq.\,(\ref{Einzel}) does not reproduce correctly an exponentially small deviation of $\Psi$ from $\Psi(0)$ at low temperatures,  it  generates there a flat behavior so that in numerical evaluation this difference may not matter.

Using $\Psi(0)/T_c$ of Eq.\,(\ref{NV}) we rewrite (\ref{Einzel}) in the form
         \begin{eqnarray}
 \frac{\Psi(t)} {\Psi(0)} =
 \tanh\left(  e^{\gamma}  \sqrt{\frac{8 (1-t) }{7\zeta(3)\,t}}\frac{
 e^{\langle\Omega^2\ln|\Omega|\rangle}  }{\sqrt{\langle\Omega^4\rangle}}   \right).\qquad
  \label{Einzel1}
\end{eqnarray}
Hence, we can evaluate the ratio $\Psi(t)/\Psi(0)=\Delta(t)/\Delta(0)$ for any particular $\Omega$.

 \begin{figure}[h]
\includegraphics[width=7cm] {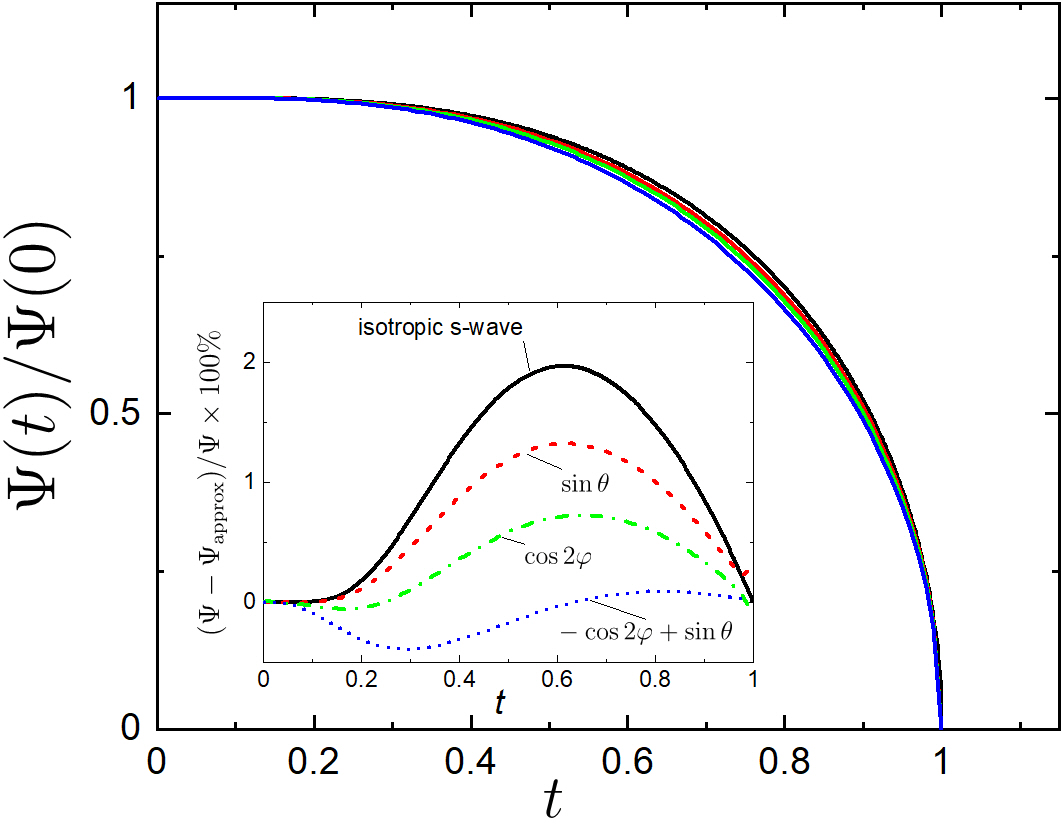}
\caption{(Color online) The order parameter $ \Psi(T)/\Psi(0) $ according to Eq.\,(\ref{Einzel1}). The black curve is the isotropic s-wave, $\Omega=1$.
The red curve is   for the clean sample with polar nodes, $\Omega = \sqrt{3/2}\sin\theta$ with $\theta$ being the polar angle on the Fermi sphere. The green curve   is for the d-wave with $\Omega = \sqrt{2}\cos 2\varphi$, and the blue one is for $\Omega \propto  (\sin \theta - \cos 2\varphi)$.  The inset shows the difference between exact  $\Psi(t)$ obtained solving numerically the self-consistency Eq.\,(\ref{self-onsc-clean}) and   interpolations (\ref{Einzel1}) for the indicated order parameters.
}
\label{f1}
\end{figure}

 The order parameter $\Delta(t)/\Delta(0)$ for   point polar nodes, normalized to its value at $T=0$, is shown in Fig.\,\ref{f1} by the red  curve. The isotropic case is shown by black for comparison, the green is for the d-wave, and the blue is for a mixed order parameter. One can say that in the chosen reduced units   all these curves overlap within a few percents accuracy. One can also say that one cannot deduce the type of order parameter from  the measured ratio  $\Delta(t)/\Delta(0)$.

\section{  $\bm{\gamma_\lambda(T )}$}

To consider the system response to a weak field, one turns to full set of Eilenberger equations:
\begin{eqnarray}
{\bm v} {\bm  \Pi}f&=&2  \Delta g   -2 \omega f   \,,\label{Eil1a}\\
 -{\bm v} {\bm  \Pi}^*f^+&=&2  \Delta^* g   -2 \omega f^+  \,,\label{Eil2a}\\
g^2&=&1-ff^{+}\,. \label{Eil3a}
 \end{eqnarray}
Here, ${\bm v}$ is the Fermi velocity, ${\bm  \Pi} =\nabla +2\pi i{\bm
A}/\phi_0$, ${\bm A}$ is the vector potential, and $\phi_0$ is the flux quantum;   $f,g$ now depend on coordinates.

Weak supercurrents and fields leave the  order parameter modulus unchanged,
but  cause the condensate to acquire an
overall phase $\chi({\bm  r})$. We therefore look for   perturbed solutions
of the Eilenberger system in the form:
\begin{eqnarray}
 \Delta\, e^{i\chi},\,\,\,\,\ (f_0   +f_1)\,e^{i\chi}, \quad
 (f_0   +f_1^+ )e^{-i\chi},\,\,\,\,\,g_0   +g_1,\qquad
\label{perturbation}
\end{eqnarray}
where $f_0,g_0$ refer to the uniform zero-field state discussed above and the subscript 1 marks corrections due to small perturbations ${\bm v} {\bm  \Pi}$.  In the London
limit, the only coordinate dependence is that of the phase $\chi$, i.e.
        $f_1 ,g_1 $ are ${\bm  r}$ independent too.

The Eilenberger equations (\ref{Eil1a})-(\ref{Eil3a}) provide the corrections among which we need only $g_1$:
\begin{equation}
g_1=\frac{i  f_0 ^2 {\bm v}{\bm P}}{2({  \Delta_0} f_0 +
      { \omega}g_0 )}= \frac{i   f_0^2 }{2 \beta_0 }
\,{\bm v}{\bm P}\,.
\label{g1}
\end{equation}
Here the super-momentum ${\bm  P}= \nabla\theta+ 2\pi{\bm  A}/\phi_0\equiv 2\pi\,
{\bm  a}/\phi_0$ with the ``gauge invariant vector potential"
 ${\bm  a}$. Substituting this in the general expression for the current density
 \begin{eqnarray}
{\bm  j}=-4\pi |e|N(0)T\,\, {\rm Im}\sum_{\omega >0}\Big\langle {\bm v}g\Big\rangle\,
\label{eil-curr}
\end{eqnarray}
and comparing the result with the London current $4\pi
j_i/c=-(\lambda^2)_{ik}^{-1}a_k$, one obtains   \cite{K2002}:
\begin{equation}
(\lambda^2)_{ik}^{-1}= \frac{16\pi^2 e^2N(0) T}{c^2}\, \sum_{\omega} \Big\langle\frac{
 \Delta^2v_iv_k}{ \beta ^{3}}\Big\rangle \,.
\label{lambda-tensor-clean}
\end{equation}

Hence, we have for the $T$ dependence of the anisotropy  $\gamma_\lambda^2(T)=\lambda_{cc}^2/\lambda_{aa}^2$ of uniaxial materials:
\begin{eqnarray}
 \gamma_\lambda^2(t)= \frac{\langle \Omega^2v_a^2 \sum_{\omega} [  \omega^2+\Psi^2(t)\Omega^2]^{-3/2}\rangle }  { \langle\Omega^2v_c^2 \sum_{\omega} [  \omega^2+\Psi^2(t)\Omega^2]^{-3/2} \rangle}\,.
 \label{gam_lam}
\end{eqnarray}
 In particular, one has:
 \begin{equation}
  \gamma_\lambda^2(0)   =\frac{  \langle v_a^2\rangle}{ \langle v_c^2\rangle} \,,\quad
    \gamma_\lambda^2(T_c)   =\frac{\langle\Omega^2   v_a^2\rangle}{ \langle \Omega^2v_c^2\rangle} \,.
\label{gamma-clean}
\end{equation}
The result for $\gamma_\lambda^2(T_c)$ is originally due to Gor'kov and Melik-Barkhudarov  \cite{Gorkov}.

Thus, the general scheme for evaluation of $\lambda(T)$   consists of two major steps: first evaluate the order
parameter $\Delta_0(T)$ in uniform zero-field state, then use  Eq.\,(\ref{lambda-tensor-clean}) with a proper averaging
over the Fermi surface. The sum over Matsubara frequencies is fast-convergent
and is  done numerically, except   limiting situations for which
analytic  evaluation is possible.

We now consider a few cases of different order parameters on a one-band Fermi sphere and show that,   depending on the order parameter, the   anisotropy  $\gamma_\lambda (T )$ might increase or decrease monotonically on warming or even be a non-monotonic function of $T$.

 \subsection{d wave}

For the d-wave $\Omega=\sqrt{2}\cos 2\varphi$.  One finds $\langle\Omega^2\ln|\Omega |\rangle=(1-\ln 2)/2$ and $\Psi(0)/T_c=\sqrt{2}\pi  e^{-0.5- \gamma} \approx 1.513$  whreas $\Delta_{max}(0)/T_c=(\Psi(0)/T_c )\sqrt{2}=2.14 $. The ratio that enters   interpolation (\ref{Einzel1}) is
        \begin{eqnarray}
\rho= \frac{e^{\langle\Omega^2\ln \Omega^2\rangle/2}}{\langle\Omega^4 \rangle}  \approx 0.78\,.
 \label{parameter}
\end{eqnarray}

 To evaluate $ \gamma_\lambda (t) $, one needs the $t$ dependence of the order parameter given in Eq.\.(\ref{Einzel1}).
 The numerical evaluation then gives  $ \gamma_\lambda (t)=  1 $ in agreement with earlier calculations of end points $ \gamma_\lambda (0)= \gamma_\lambda (T_c)=1 $ \cite{Kosh}.

 \subsection{Polar nodes on Fermi sphere}

 We model this case by setting $\Omega=\sqrt{3/2}\sin\theta$. One readily finds $\gamma_\lambda^2(T_c)=2$.
 Further, we obtain
        \begin{eqnarray}
 \frac{1}{2}\langle\Omega^2\ln \Omega^2\rangle = \frac{\ln 216-5}{6}\approx 0.0626.
 \label{Om ln Om sin}
\end{eqnarray}
and the parameter $\rho\approx 0.89$.
 \begin{figure}[h]
 \includegraphics[width=6.5cm] {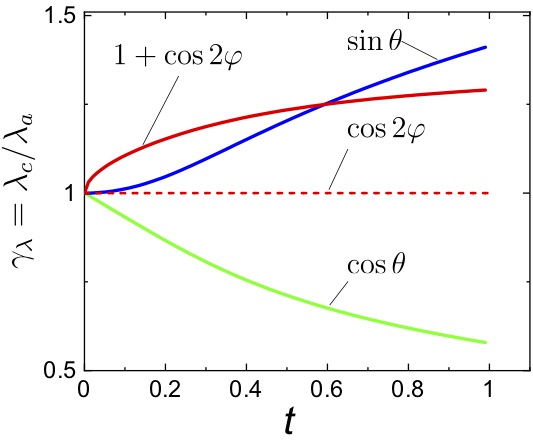}
\caption{(Color online)  Anisotropy parameters $\gamma_\lambda$ vs the reduced temperature $t$ for the order parameter with polar point nodes $ \Omega=\sqrt{3/2}\sin\theta  $ (blue), the equatorial line node $\sqrt{3 }\cos\theta  $ (green), and  the d-wave $\sqrt{ 2}\cos 2\varphi$ (dashed red). The curve $1+\cos 2\varphi$ for a mixed $s$ and $d$ order parameters increases with $t$, despite the fact that $s$ and $d$ separately have $t$ independent $\gamma_\lambda=1$.
}
\label{f2}
\end{figure}
The anisotropy parameter evaluated numerically as described above is shown by the blue curve in Fig.\,\ref{f2} which shows that $\gamma_\lambda(t)$ increases. If $\Omega\propto \cos\theta$ the same numerical procedure yields the decreasing $\gamma_\lambda(t)$. Interestingly, a pure d-wave order parameter $\Omega=\sqrt{2} \cos2\varphi$ as well as pure s-wave $\Omega=1$ produce a temperature independent $\gamma_\lambda(t)=1$, whereas their mixture, e.g. $1+\cos2\varphi$, gives an increasing  $\gamma_\lambda(t)$.

To check accuracy of Einzel's approximation for $\Psi(T)$
we did all calculations based on Eilenberger theory {\it per se} and we find no noticible differences.

 \subsection{Equatorial line node }

This type of line node was suggested as possible in some Fe based materials \cite{vivek-Hirschfeld,equat-node} and observed in ARPES experiments \cite{equat-node2}.
For the order parameter $\Omega=\sqrt{3}\cos\theta$ we evaluate:
        \begin{eqnarray}
 \frac{1}{2}\langle\Omega^2\ln \Omega^2\rangle = \frac{\ln 27-2}{6}\approx 0.216
 \label{Om ln Om cos}
\end{eqnarray}
and the parameter $\rho\approx 0.69\,$.
The corresponding $\gamma_\lambda(t)$ is shown by the green curve in Fig.\,\ref{f2}. Thus, on the basis of this and the previous example one concludes that, depending on the order parameter, $\gamma_\lambda $ may increase or decrease on warming even in one band systems.

\subsection{$\bm{ \Omega=\Omega_0 (a+ \cos{\bm 2 \varphi})} $}

This corresponds to a mixed s and d-wave order parameter, a possibility considered for cuprates, see e.g. \cite{Openov,Nagi}.

\begin{figure}[h]
\includegraphics[width=8cm] {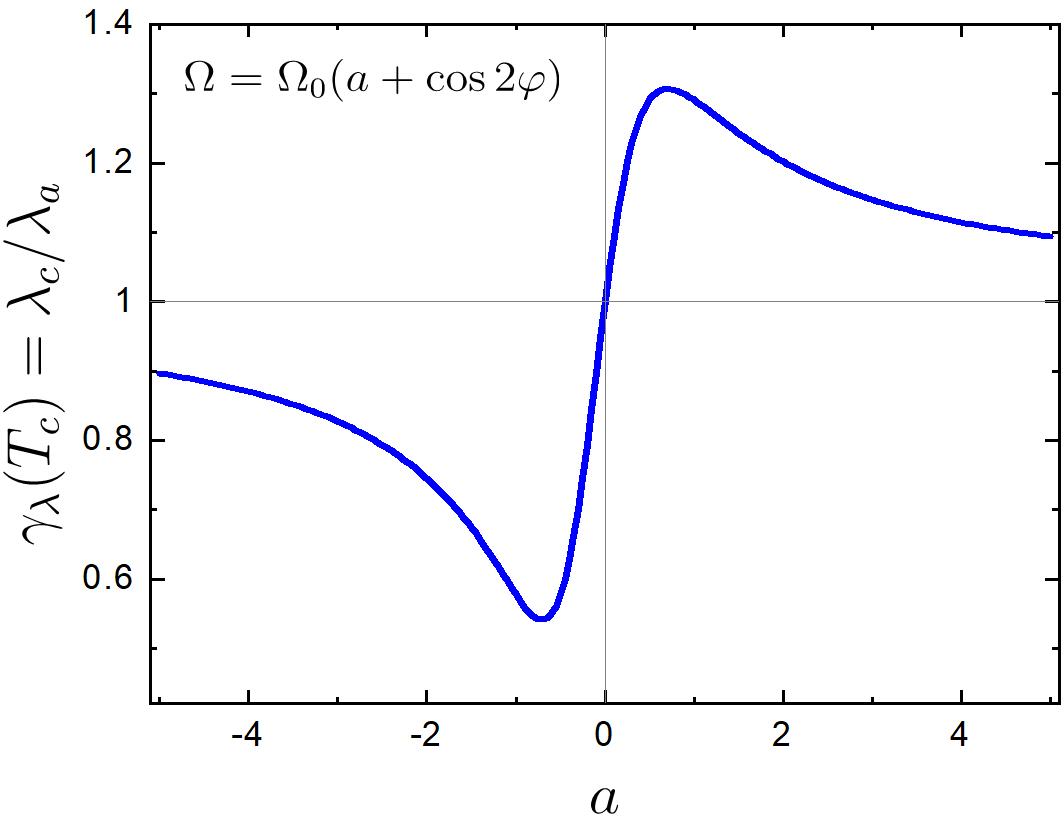}
\caption{(Color online)  $\gamma_\lambda(T_c) $ vs $a$ for the order parameter $\Omega=\Omega_0(a+\cos 2\varphi)$.    }
\label{ff10}
\end{figure}

The anisotropy parameter $\gamma_\lambda (T_c) $ vs $a$ is shown in Figs.\,\ref{ff10}. Since on a Fermi sphere  $\gamma_\lambda^2(0) =1$, one sees that for $a>0$ the anisotropy $\gamma_\lambda (T ) $ grows on warming, whereas for negative $a$ it decreases. Surprising at first sight, this means that for  mixed order parameters  $\gamma_\lambda (T ) $ depends on  relative phases of the order parameters in the mixture; in this case for $a<0$ the phase difference is $\pi$.

\begin{figure}[h]
\includegraphics[width=8cm] {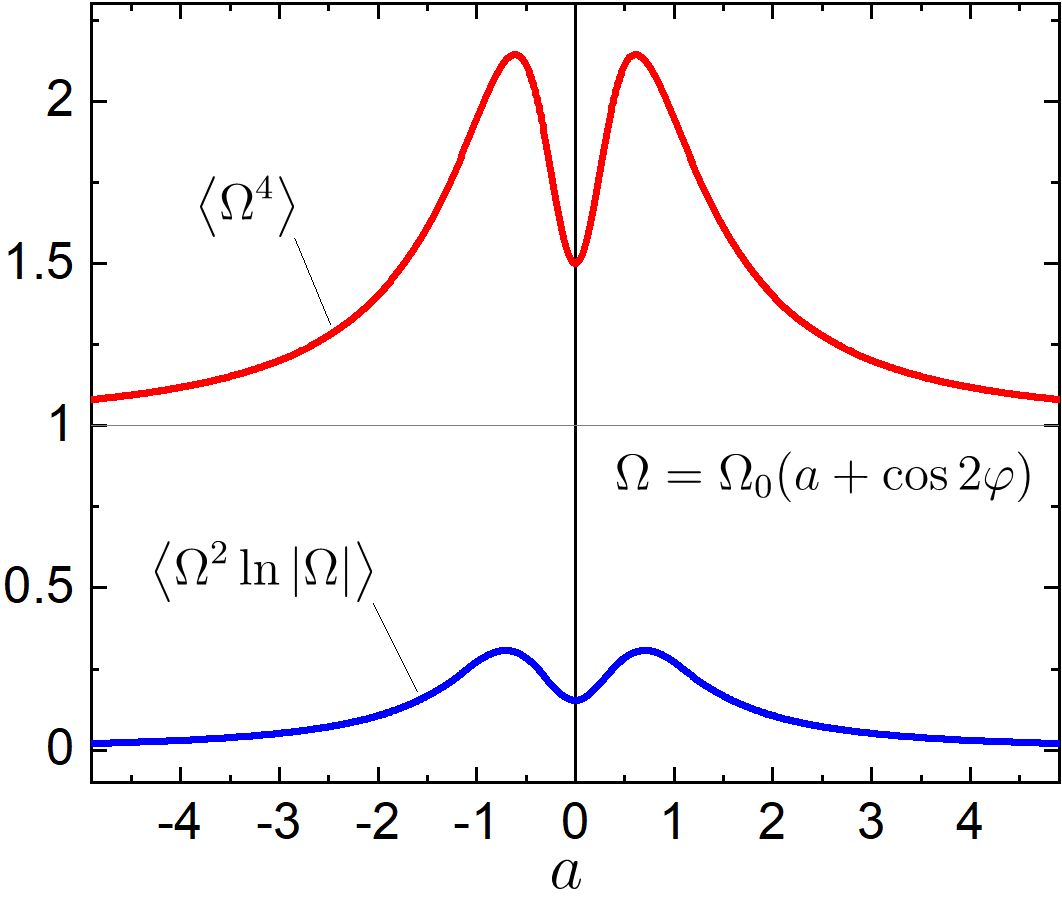}
\caption{(Color online) The top curve  of $\langle\Omega^4\rangle$ vs $a$ and the low curve is  $\langle\Omega^2\ln |\Omega |\rangle$ for the order parameter $\Omega=\Omega_0(a+ \cos 2\varphi)$.    }
\label{f4}
\end{figure}

The upper curve in Fig.\,\ref{f4} shows the parameter $\langle\Omega^4\rangle$ which affects  the specific heat jump \cite{Openov,Nagi,Einzel}
 \begin{eqnarray}
  \frac{\Delta C}{C_n(T_c)} = \frac{12}{7\zeta(3)\langle \Omega ^4\rangle} \approx \frac{1.426}{ \langle \Omega ^4\rangle}\,.
  \label{Cjump}
\end{eqnarray}
The lower curve is the parameter $\langle\Omega^2\ln |\Omega |\rangle$ which enters the ratio $\Psi(0)/T_c $, Eq.\,(\ref{NV}). Since this parameter is small at all $a$ one has
\begin{eqnarray}
 \Psi(0)/T_c \approx 1.76 \,(1-\langle\Omega^2\ln |\Omega |\rangle)\,.
  \label{<del/Tc(s+d)>}
\end{eqnarray}
In fact, $\langle\Omega^2\ln |\Omega |\rangle <1$ in all examples we have considered.

\subsection{$\bm{ \Omega=\Omega_0 (a+\sin{\bm\theta})} $}

\begin{figure}[h]
\includegraphics[width=6cm] {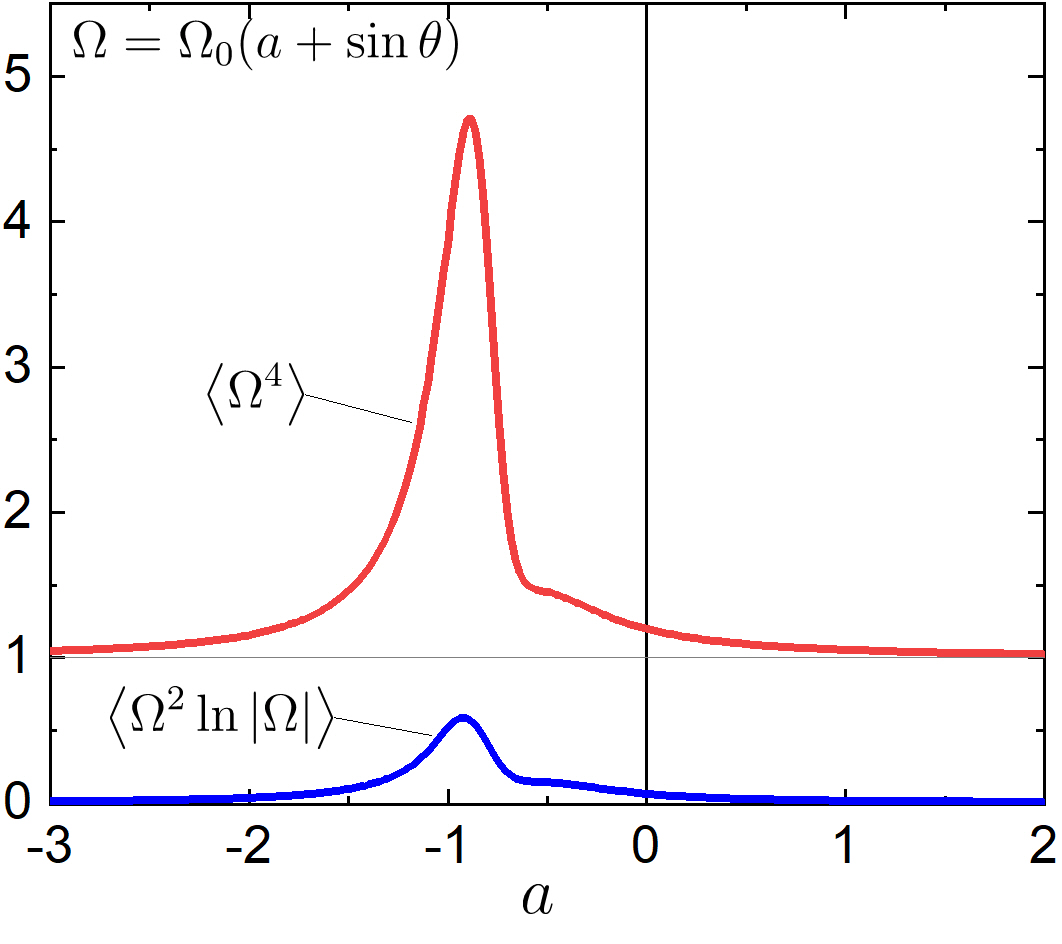}
\caption{(Color online) $ \langle \Omega ^4\rangle$ and $\Omega^2\ln |\Omega |$ vs $a$ for the order parameter $ \Omega=\Omega_0 (a+\sin\theta) $. }
\label{f5}
\end{figure}

This corresponds to a mixture of s-wave and the phase with polar nodes. It is instructive to study this case, because positive $a$'s make the condensate a nodeless anisotropic s-wave, whereas $a<0$ turns the polar nodes into line nodes along certain altitude circles. We start with the normalization $\langle\Omega^2\rangle=1$ which yields
 \begin{eqnarray}
  \Omega_0^2 = \frac{2}{4/3+\pi a+2a^2} \,.
  \label{<Om0>}
\end{eqnarray}
Next we calculate
\begin{eqnarray}
 \langle \Omega ^4\rangle &=& \frac{4}{(4/3+\pi a+2a^2)^2}  \langle ( a+\sin\theta)^4\rangle\nonumber\\
&=&2 \frac{16/15+8a^2+2a^4+3\pi a/2+2\pi a^2}{(4/3+\pi a+2a^2)^2} \,.\qquad
  \label{<Om4(a)>}
\end{eqnarray}
This function is plotted in Fig.\,\ref{f5}.
The maximum of this curve at $a_m\approx -0.888$ means that  the order parameter near $T_c$ of Eq.\,(\ref{GL-OP}) along with the specific heat jump
are suppressed at  $a_m$ by about a factor of 5 relative to pure s-wave.

It is instructive also to plot the ratio $\Delta(0)/T_c$ which is traditionally considered as distinguishing parameter for  weak and strong couplings. If the order parameter is anisotropic, $\Delta(0)_{max}/T_c$ is usually measured. Using Eq.\,(\ref{NV}) we obtain
         \begin{eqnarray}
\frac{\Delta(0;a,\theta)}{T_c}= \frac{\Psi(0)\Omega(a,\theta)} {T_c} = \frac{\pi \Omega } { e^{ \gamma}}\,e^{-\langle\Omega^2\ln|\Omega |\rangle}.
 \label{ratio}
\end{eqnarray}
The absolute value of this ratio as a function of $a$ is plotted for $\theta=0$ and $\pi/2$ in Fig.\,\ref{f6}:
   \begin{figure}[h]
\includegraphics[width=7cm] {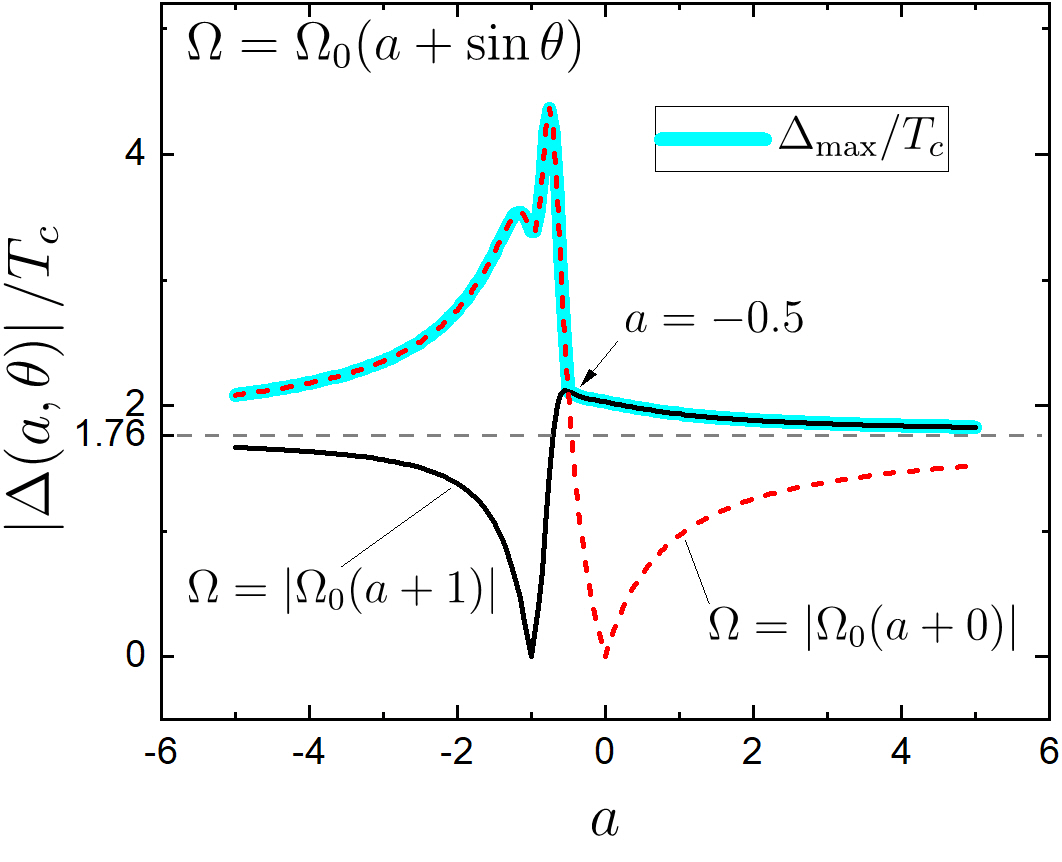}
\caption{(Color online)  $|\Delta_{max}(0)|/T_c $ vs $a$ for  $ \theta=0$ (red dashed curve) and $\theta= \pi/2$ (black curve).  }
\label{f6}
\end{figure}

\begin{figure}[h]
\includegraphics[width=7cm] {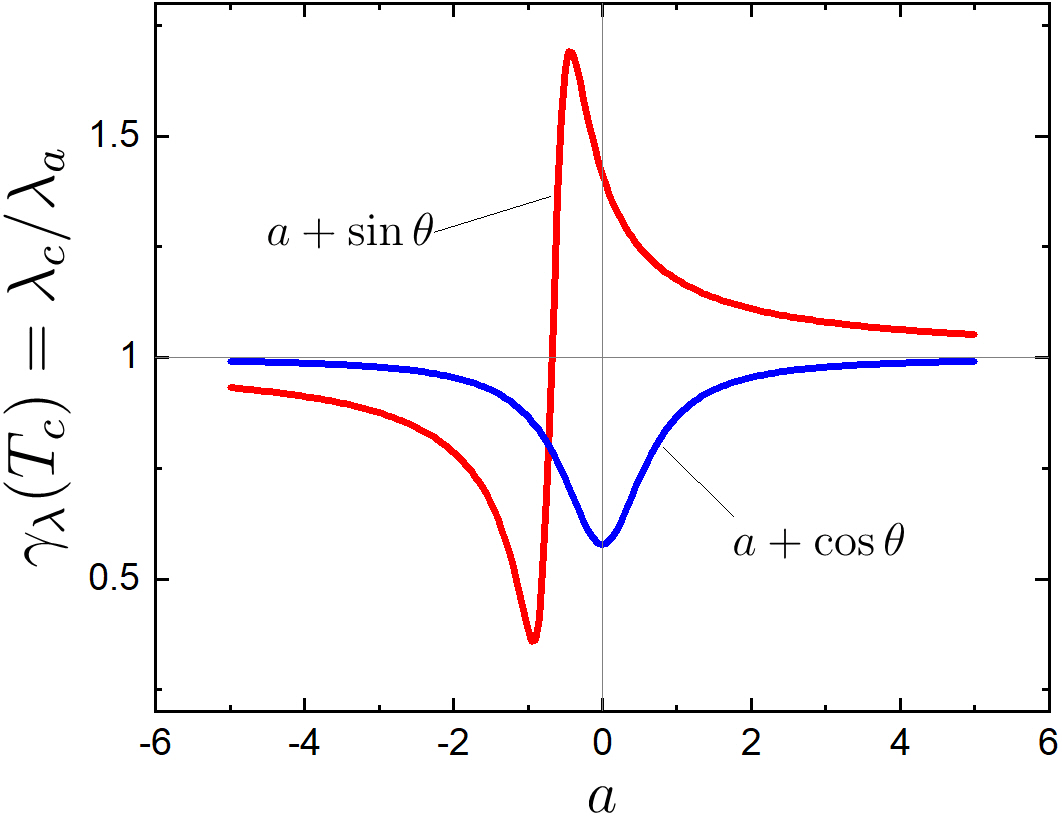}
\caption{(Color online) $\gamma_\lambda^2(T_c) $ vs $a$ for the order parameter $\Omega=\Omega_0(a+\sin\theta)$.  $\gamma_\lambda^2(T_c) =1$ at $a=-0.68$.}
\label{f7}
\end{figure}

 After straightforward algebra one   obtains for the anisotropy of penetration depth at $T_c$:
\begin{eqnarray}
  \gamma_\lambda^2(T_c)   = \frac{\langle\Omega^2   v_a^2\rangle}{ \langle \Omega^2v_c^2\rangle} =   \frac{64+ 45\pi a +80 a^2}{120(4/15+\pi a/4+2a^2/3) } \,.\qquad
  \label{<g2tc)>}
\end{eqnarray}
  This function is plotted in Fig.\,\ref{f7}.  The reason for the asymmetry of this plot relative to $a=0$ is clear: for $a>0$ the polar nodes are no longer exist and the phase becomes an anisotropic $s$. A similar
situation takes place for $a\lesssim -1$ where the $s$ part acquires a minus sign (or an extra phase shift of $\pi$). The most interesting part corresponds to the sharp drop of the curve in the interval $-1\lesssim a\lesssim -0.5$, where the  point polar nodes  transform to line circular nodes on the altitude $\theta=-\arcsin a$.

  Since on the Fermi sphere $\gamma_\lambda(0)=1$, this figure
gives an idea of how  $\gamma_\lambda(T)$ may behave when the temperature varies from 0 to $T_c$. One can see that for $a>-0.68$, where the curve of $\gamma_\lambda(t) $ (shown in red) crosses the line $\gamma_\lambda =1  $, i.e. in anisotropic nodeless  $s$ phase $\gamma_\lambda(0)<\gamma_\lambda(T_c ) $.
  \begin{figure}[h]
\includegraphics[width=7cm] {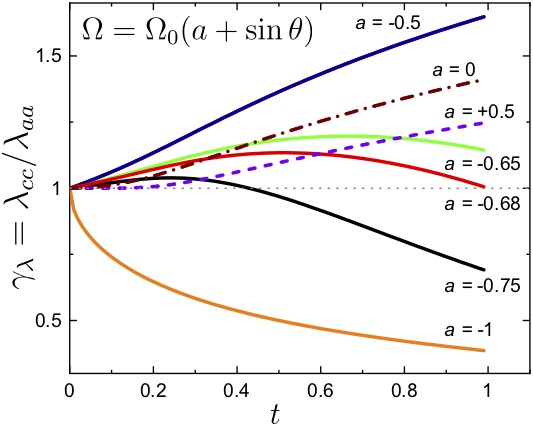}
\caption{ (Color online) $\gamma_\lambda  $ vs $t$ for the order parameter $\Omega=\Omega_0(a+\sin\theta)$ for a set of $a $'s indicated.
}
\label{f8}
\end{figure}

In a relatively narrow interval of values of the parameter $a$ near $a \approx - 0.7$, $(\gamma_\lambda -1)$ changes fast from positive  to  negative values, i.e. from increasing $\gamma_\lambda(t) $ to decreasing. The question then arises whether in this transformation domain $\gamma_\lambda(t) $ remains monotonic. Examples in   Fig.\,\ref{f8} for $a=-0.75, -0.68$ and -0.65 show  that this is not the case, $\gamma_\lambda(t) $ clearly has a well pronounced maximum.
This   figure demonstrates the evolution of  the shape of  $\gamma_\lambda(t) $ with changing weigh $a$ of the s-wave fraction in the order parameter $\Omega=\Omega_0 (a+\sin{\theta})$.

Thus,   depending on the relative weight of two phases involved, we can have $\gamma_\lambda $ increasing or decreasing on warming, the features commonly associated with multi-gap superconductivity.

\subsection{$\bm{ \Omega=\Omega_0 (a\cos 2\varphi +\sin{\bm\theta})} $}

This mixture of d-wave order parameter with line nodes at two meridians on the Fermi sphere and the polar point nodes differs from the previous example because polar nodes remain in the presence of d-wave, whereas line nodes do not survive due to term $\sin\theta$.
The treatment of this situation is similar to the cases considered, so that  we show  only the results.
  \begin{figure}[htb]
\includegraphics[width=8cm] {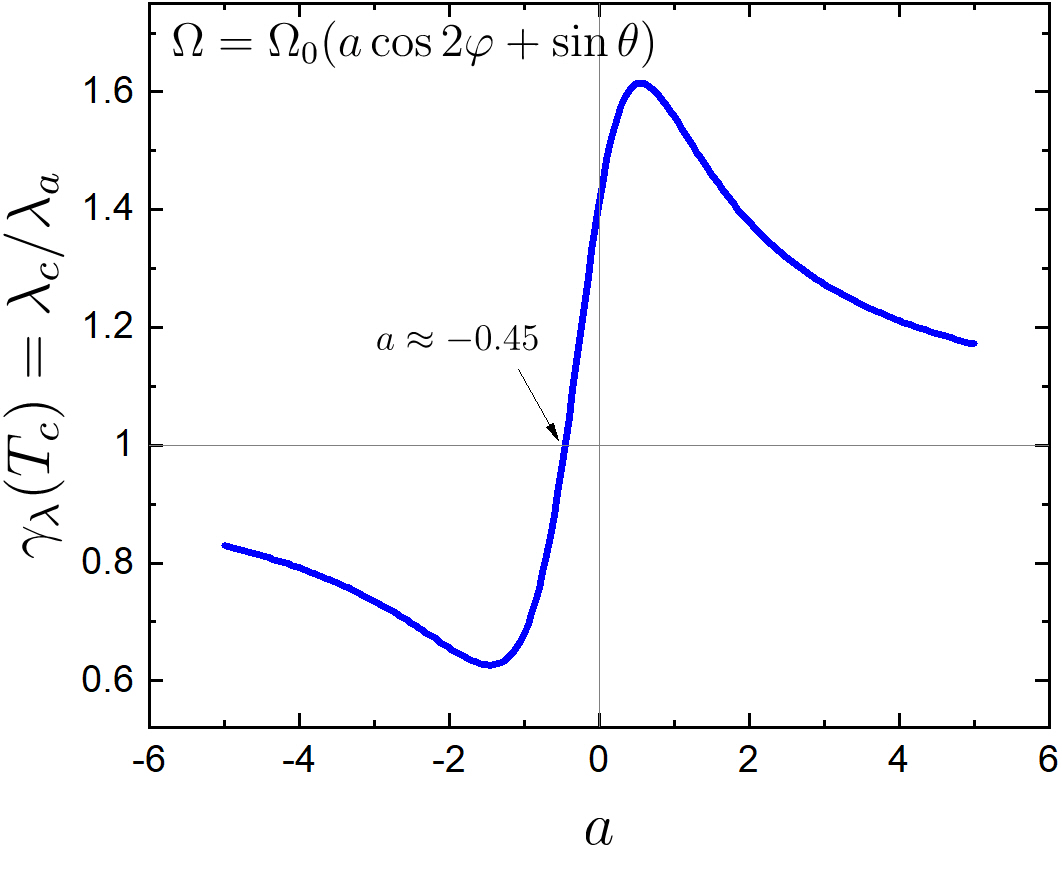}
\caption{(Color online) $\gamma_\lambda(T_c)$ vs $a$ for the order parameter $\Omega=\Omega_0(a\cos 2\varphi+\sin\theta)$.
}
\label{f9}
\end{figure}

\begin{figure}[htb]
\includegraphics[width=7.5cm] {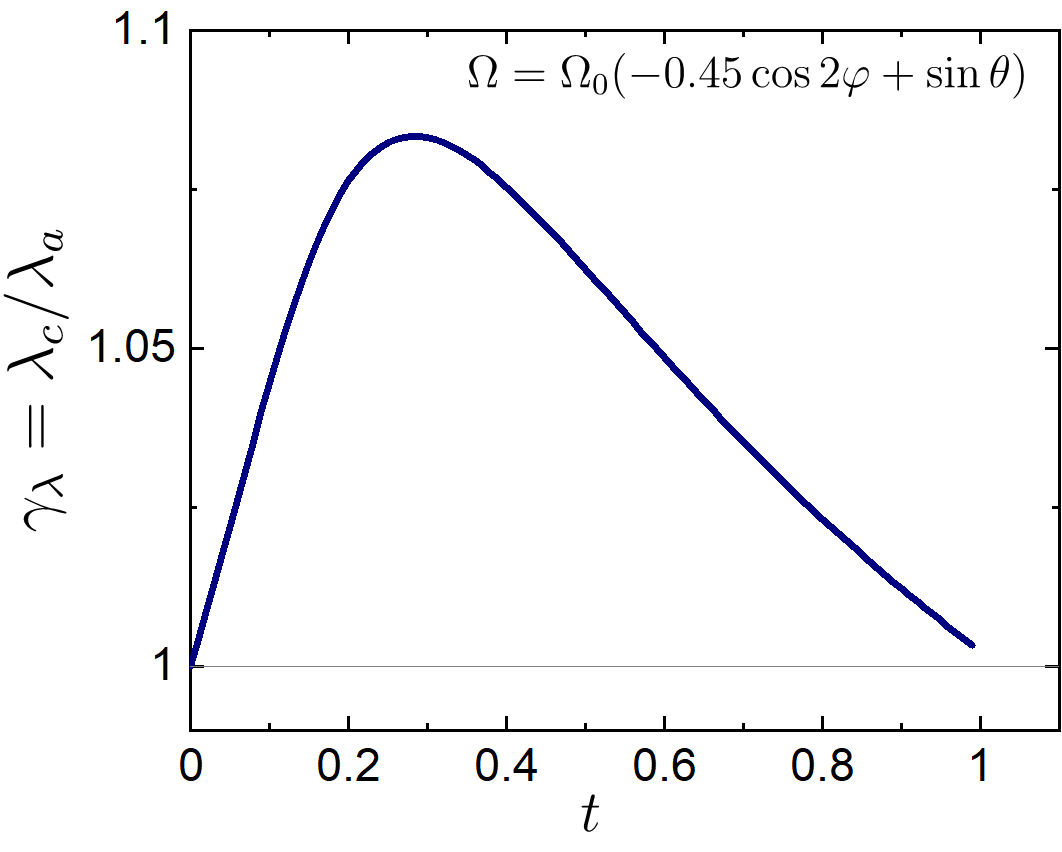}
\caption{(Color online) $\gamma_\lambda$ vs $t$ for the order parameter $\Omega=\Omega_0(a\cos 2\varphi+\sin\theta)$ with $a=-0.45$.
}
\label{f10}
\end{figure}

The anisotropy parameter $\gamma_\lambda(T_c)$ for this case is shown
 in Fig.\,\ref{f9}. A sharp drop in the interval $-1\lesssim a\lesssim 0.4$ reminds a similar drop for $\Omega=\Omega_0(a+\sin\theta)$,  the mixture of s-wave and polar nodes.
We expect a non-monotonic $\gamma_\lambda(t)$ in the vicinity of $a\approx -0.4$ where $\gamma_\lambda^2(T_c)-1$ changes sign. Indeed, we see this in  Fig.\,\ref{f10}. Hence, the maximum of
$\gamma_\lambda(t)$ which we found for   another mixed order parameter $\Omega=\Omega_0(a+\sin\theta)$, Fig.\,\ref{f8}, was not accidental.

\subsection{On  ratio of  experimental energy gap to $\bm{T_{c}}$}

The ratio $R=\Delta(0)/T_{c}$ is one of the
fundamental  superconducting parameters  that can be measured experimentally. However, there is a
great deal of confusion in  experimental literature as to what shall one
expect within the weak-coupling BCS theory (which differs from the "strong coupling" Eliashberg approach). Often this ratio, determined
from  spectroscopic measurements (STM, ARPES, optical reflectivity), is   larger than that
determined from  thermodynamic experiments (the thermodynamic critical field$ H_c$, the specific heat jump at $T_c$, the superfluid density). We have shown, however, that this ratio may exceed the BCS prediction of $R\approx 1.76$ within a weak-coupling BCS models for anisotropic order parameters. Hence,  the measured $R> 1.76$   might not serve as evidence for strong coupling.

The energy gap that enters the thermodynamics cannot exceed the isotropic
s-wave BCS value of $\Delta(0)/T_{c}\approx1.76$. Specifically, one can
measure $H_{c}$, the specific heat jump at
$T_{c}$ or the superfluid density to determine this gap. The condensation energy at $T=0$
\begin{eqnarray}
F(0)   =\frac{H_{c}^{2}}{8\pi}=\frac{N(0)}{2}\langle\Delta^2(0)\rangle= \frac{N(0)}{2}\Psi^{2}(0)\,,
\label{F(0)}
\end{eqnarray}
that gives $\Psi (0)=H_c(0)/\sqrt{4\pi N(0)}$.
According to Eq.\,(\ref{NV})
$\Psi(0)/  T_{c}  =1.76\, \exp(-\langle   \Omega^{2}\ln\Omega^{2}\rangle/2 )$.
In all cases we have studied $0< \langle
\Omega^{2}\ln\Omega^{2} \rangle<1$   so that $\Psi(0)/T_{c}$ does not exceed the weak coupling value of 1.76. Hence, if one extracts the gap from the data on   $H_c$, the ratio $R$ is expected to be less than 1.76. Also,
measurements of the superfluid density \cite{Prozorov-Kogan-RPP} provide the  magnitude of the order parameter $\Psi(T)$.

The specific heat jump is given in Eq.\,(\ref{Cjump}).
In all cases we have considered  $\left\langle \Omega^{4}\right\rangle \geq1,$ see Figs.\,\ref{f4} and \ref{f5} so that the jump is smaller than the isotropic value of 1.43.

\begin{figure}[h]
\includegraphics[width=7.5cm] {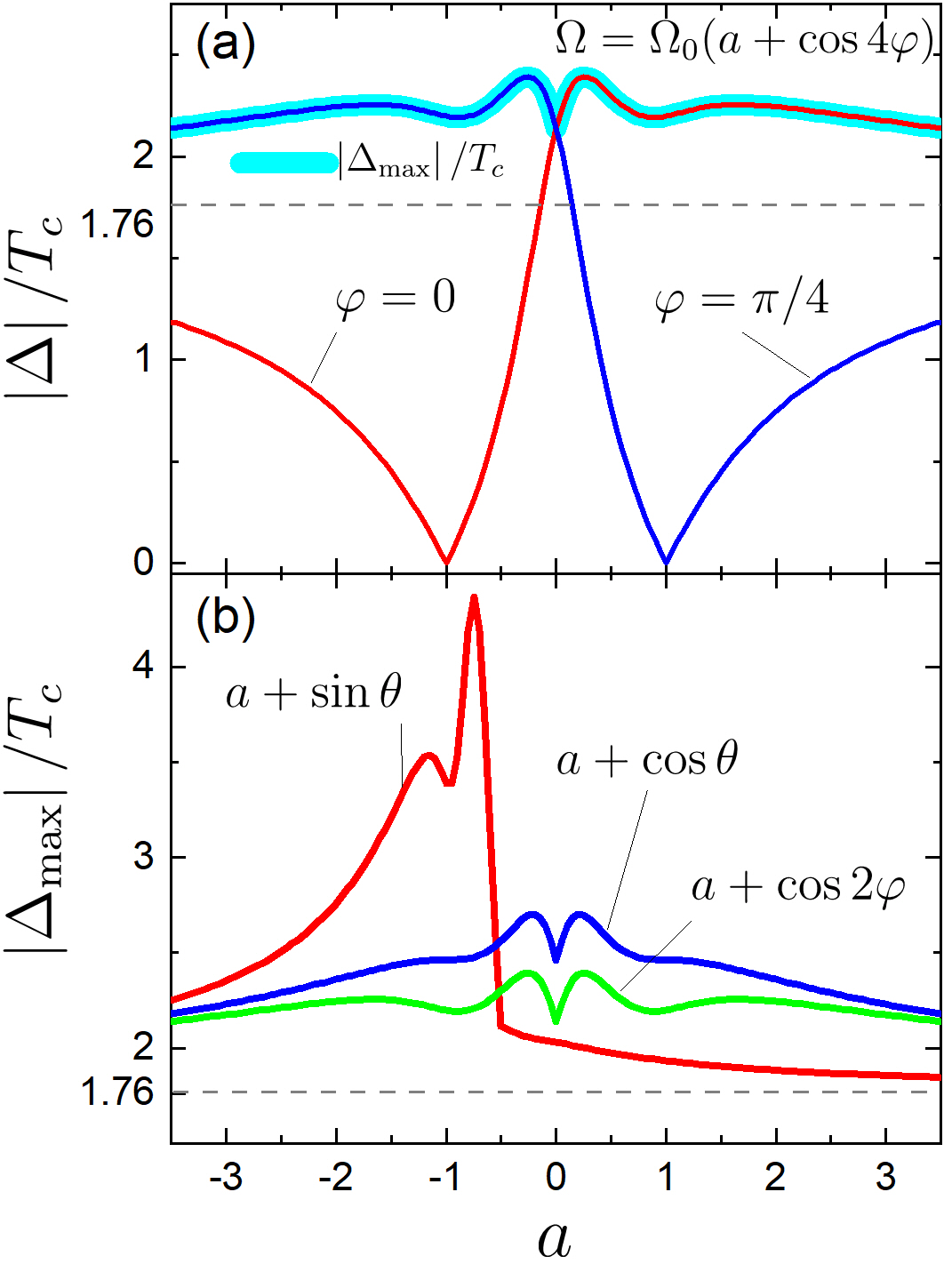}
\caption{(Color online) The upper panel: $\Delta(0)/T_c$ vs $a$, the weight parameter of ad-mixture s-phase, for order parameters $  \Omega=\Omega_0 (a  +\cos 4 \varphi)$ which describe ARPES data for KFe$_2$AS$_2$ \cite{Fe-based}. The lower panel: $|\Delta_{max}(0)|/T_c$ vs $a$  for order parameters indicated.
}
\label{f11}
\end{figure}

The spectroscopic gaps (actually, the gaps in the quasiparticle spectrum) determined in ARPES, optical reflectivity, and tunneling
experiments are a different story. Here,  experiments give s the
maximum value of the superconducting gap:
 \begin{eqnarray}
\Delta_{\max}(0)=| \Psi(0)\Omega_{\max} ( \bm k ) |\,.
\end{eqnarray}
The normalization $\langle \Omega^2 \rangle=1$ implies that $\Omega_{\max} \geq 1$, i.e., $\Delta_{\max}(0)\geq \Psi(0)$.
 It is shown  in Figs.\,\ref{f6} and \ref{f11} that, indeed, the ratio $\Delta_{\max}(0)/T_c$ differs
from the thermodynamic ratio $\Psi(0)/T_c$.

To conclude, the ``thermodynamic" gap ratio is   less or equal to the isotropic weak-coupling BCS value of $1.76$ whereas the maximum gap from spectroscopic experiments over
$T_{c}$ is   greater than that. This difference
led to often erroneous assignment of the
larger than BCS values to the strong coupling. But the arguments we present here  are developed, in fact, on the basis of weal coupling Eilenberger theory for  anisotropic order parameters.

These arguments can be extended
to  multi-band systems. Specifically, within the weak-coupling model of  two-band superconductors, one gap
will always be greater and the other   smaller than the BCS value.

Thus, experimental ratios   $\Delta_{\max}(0)/T_c$ cannot be used to claim strong coupling without knowledge of the order parameter anisotropy. On the
other hand, comparative analysis of thermodynamic and spectroscopic gaps may
be used if not to determine, but definitely to restrict the possible order
parameters for a particular material.

 \section{Discussion}

 The separable coupling model for one band Fermi surfaces not only reproduces weak coupling isotropic BCS thermodynamics, but allows one to incorporate  anisotropies of Fermi surfaces and of condensate order parameters. In particular, it provides a relatively straightforward procedure to obtain the temperature dependence of penetration depth and its anisotropy. As is the case in BCS, this procedure involves determination of the equilibrium order parameter $\Delta(T)$ by solving the self-consistency equation   (the gap equation), a ``labor intensive" part in anisotropic case. An alternative approach was  given in Refs.\,\cite{Peter,Peter2,Einzel} where an accurate  analytic interpolation for $\Delta(T)$ was   offered that could be used instead of solving the self-consistency equation.
  We have veryfied this procedure for a number of different order parameters by comparing with the numerical solutions of the self-consistency equation and we found only small differences in the results insignificant as far as the accuracy of existing experimental data is concerned.

 To separate possible effects of the order parameter anisotropy from those of Fermi surfaces, we considered only the   Fermi sphere.
 We found that the anisotropy parameter of the penetration depth increases on warming for the order parameter with point nodes at the poles of the Fermi sphere, $\Omega=\sqrt{3/2}\sin\theta$. However, for the order parameter with a line node on the equator, $\Omega=\sqrt{3}\cos\theta$, $\gamma_\lambda(t)$ decreases. We have confirmed that for the d-wave, $\Omega=\sqrt{2}\cos(2\varphi) $, $\gamma_\lambda(T)=1$ at all temperatures in agreement with previously calculated end point values  $\gamma_\lambda(0)=\gamma_\lambda(T_c)=1$ \cite{Kosh}. Thus, a common way to attribute the $T$ dependence of $\gamma_\lambda(T) $ to different gaps at multi-band Fermi surfaces  is clearly questionable.

The possibility of mixture of order parameters of different symmetries has been discussed for cuprates  and other superconductors, see e.g.\,\cite{Nagi,Openov, Fe-based}. Our analysis of the order parameter $\Omega=\Omega_0(a+\cos\,2\varphi)$ showed that the anisotropy $\gamma_\lambda(T) $ depends on
the relative phase of the constitutive order parameters ($\pi$ for $a<0$).

 We have considered  $\Omega=\Omega_0(a+\sin\theta)$, where $a$ is the relative weight of the s-wave phase as compared to the order parameter with polar nodes.  First, we find that the ratio $\Delta_{max}(0)/T_c$ may exceed considerably the standard weak coupling value of 1.76 in a certain region  of the parameter $a$, see Fig.\,\ref{f4}. Second, it turned out that $\gamma_\lambda(T) $ may monotonically increase or decrease and even go through a maximum depending on the relative weight $a$ of two order parameters involved.

  We have tested also the order parameter $  \Omega=\Omega_0 (a\cos 2\varphi +\sin \theta )$, i.e. a mixture of d-wave   with the phase having polar nodes. Again, we see maximum in $\gamma_\lambda(t)$  for the weight $a$ near the value which corresponds to the end values $\gamma_\lambda(0)=\gamma_\lambda(T_c)\approx 1$, Fig.\,\ref{f10}. We speculate that if experiment shows a non-monotonic anisotropy of $\lambda$, the likely reason is   a mixed order parameter.
  The last feature is intriguing in particular, because we have an experimental example of SrPt$_3$P in which  $\gamma_\lambda(T) $ goes through a maximum \cite{Kyuil}.

 As a bi-product of our results we show in Fig.\,\ref{f11} the ratio $|\Delta(0)|/T_c$ vs  the weight $a$ of ad-mixture s-phase for the order parameter  $  \Omega=\Omega_0 (a  +\cos 4\varphi )$ (the candidate for KFe$_2$As$_2$ \cite{Fe-based}) for $\varphi =0$ and $\pi/4$. It is worth noting that this ratio  differs from the isotropic weak coupling BCS $\pi e^{-\gamma}=1.76$; in fact, this ratio at certain ad-mixtures  of s-wave phase can be bigger or smaller than the BCS number. This, however, does not mean the coupling in these case is strong or it is ``weaker than weak", rather it is caused by the order parameter anisotropy. Note that experimentally measured ratio is usually $|\Delta_{max}(0)|/T_c=|\Psi(0)|\Omega_{max}/T_c$

\section{Acknowledgements}
The authors are grateful to Peter Hirschfeld for many useful, informative, and critical discussions.
The work was supported by the U.S. Department of Energy (DOE), Office of Science, Basic Energy Sciences, Materials Science and Engineering Division.  Ames Laboratory  is operated for the U.S. DOE by Iowa State University under contract \# DE-AC02-07CH11358.

\appendix

\section{Clean case order parameter at $\bm {T=0}$}

Commonly, the effective coupling $V$   is assumed factorizable
\cite{MarkKad}, $ V({\bm v},{\bm v}^{\prime\,})=V_0 \,\Omega({\bm v})\,\Omega({\bm v}^{\prime\,})$. One then looks for the order parameter in the form  $\Delta (
{\bm  r},T;{\bm v})=\Psi ({\bm  r},T)\, \Omega({\bm v})$. The coupling constant $V_0$ is chosen to get the isotropic BCS result for $\Omega=1$:
 \begin{eqnarray}
 \frac{1}{N(0)V_0}=\ln \frac{2 \omega_D}{\pi T_ce^{-\gamma}} \,,
\label{BCS}
\end{eqnarray}
 $ \omega_D$ is the energy scale of the ``glue" excitations (of phonons in conventional materials), and $\gamma\approx 0.577$ is the Euler constant.

The self-consistency  equation can be written in the form:
       \begin{equation}
\Psi( {\bm  r},T)=2\pi T N(0)V_0 \sum_{\omega >0}^{\omega_D} \Big\langle
\Omega({\bm v} ) f({\bm v} ,{\bm  r},\omega)\Big\rangle \,.
\label{gap}
\end{equation}

 Since in the clean case
$f  = \Delta  /\sqrt{\Delta ^2+  \omega^2}$  we have at $T=0$:
        \begin{eqnarray}
 \frac{1} {N(0)V_0} &=&2\pi T  \sum_{\omega >0}^{\omega_D} \left\langle\frac{
\Omega^2}{\sqrt{\Delta ^2+  \omega^2}} \right\rangle \nonumber\\
&=& \left\langle\Omega^2\int_0^{\omega_D}\frac{d
\omega }{\sqrt{\Delta ^2+ \omega^2}} \right\rangle = \left\langle\Omega^2 \ln\frac{2
\omega_D }{|\Delta|} \right\rangle.\qquad
\label{T=0}
\end{eqnarray}
 Hence, as follows from (\ref{T=0}) and (\ref{BCS}) \cite{Einzel,anis-criteria}:
         \begin{eqnarray}
 \frac{\Psi(0)} {T_c} = \frac{\pi} { e^{ \gamma}}\,e^{-\langle\Omega^2\ln|\Omega |\rangle}.
 \label{NVb}
\end{eqnarray}
Clearly, for s-wave gaps, $\Omega=1$ (at any Fermi surface) this gives  $\Delta(0)/T_c=\pi  e^{- \gamma}\approx 1.76$.

\end{document}